\documentclass[prl,aps,reprint,superscriptaddress,showpacs,twocolumn,floatfix]{revtex4-1}

\usepackage{soul}
\usepackage{ulem}

\usepackage{graphicx}
\usepackage[ansinew]{inputenc}
\usepackage{array}
\usepackage{color}
\usepackage{amsmath}
\usepackage{amsxtra}
\usepackage{amstext}
\usepackage{amssymb}
\usepackage{latexsym}
\begin{document}

\title{Long wavelength local density of states oscillations near graphene step edges}

\author{Jiamin Xue}
\affiliation{Department of Physics, University of Arizona, Tucson, AZ 85721 USA.}
\author{Javier Sanchez-Yamagishi}
\affiliation{Department of Physics, Massachusetts Institute of Technology, Cambridge, MA 02138 USA.}
\author{K. Watanabe}
\author{T. Taniguchi}
\affiliation{Advanced Materials Laboratory, National Institute for Materials Science, 1-1 Namiki, Tsukuba 305-0044, Japan.}
\author{Pablo Jarillo-Herrero}
\affiliation{Department of Physics, Massachusetts Institute of Technology, Cambridge, MA 02138 USA.}
\author{Brian J. LeRoy}
\email{leroy@physics.arizona.edu}
\affiliation{Department of Physics, University of Arizona, Tucson, AZ 85721 USA.}

\date{\today}

\begin{abstract}
Using scanning tunneling microscopy and spectroscopy, we have studied the local density of states (LDOS) of graphene over step edges in boron nitride. Long wavelength oscillations in the LDOS are observed with maxima parallel to the step edge.  Their wavelength and amplitude are controlled by the energy of the quasiparticles allowing a direct probe of the graphene dispersion relation.  We also observe a faster decay of the LDOS oscillations away from the step edge than in conventional metals.  This is due to the chiral nature of the Dirac fermions in graphene.
\end{abstract}


\maketitle

Defects, impurities and edges have a large effect of the density of states in metals and semiconductors.  Their spatial extent is usually limited by screening which suppresses the long-range behavior of the potential.  However, at the fermi wavevector, $k_F$, the electron gas has a singularity leading to strong and long range oscillations in the electron density known as Friedel oscillations.  Friedel oscillations have been observed near step edges on the surface of many metals such as Cu\cite{Crommie:1993co} and Au\cite{Hasegawa:1993dq}.  These materials are described by Fermi liquid theory and the oscillations have a characteristic wavelength given by twice the Fermi wavevector and a slow power law decay with distance.  Graphene is a two-dimensional material where the quasiparticles obey the Dirac equation and therefore behave as massless Dirac fermions\cite{Neto:2009cl}.  Many unusual effects ranging from a different quantization of the quantum Hall effect\cite{Novoselov:2005es,Zhang:2005gp} to the suppression of backscattering\cite{Neto:2009cl} have been observed in graphene.  These effects are due to the chiral nature of the quasiparticles in graphene as well as the linear dispersion relation.  These two properties also affect the Friedel oscillations and are expected to cause a faster decay with distance\cite{Cheianov:2006ku}.

There have been several scanning tunneling spectroscopy studies of graphene on SiC\cite{Rutter:2007epa,Brihuega:2008ip} and SiO$_2$\cite{Zhang:2009ce,Deshpande:2009hx} which have shown interference patterns due to quasiparticle scattering.  These studies have observed standing wave patterns due to scattering from defects in both monolayer and bilayer graphene and verified that the chiral nature of the quasiparticles leads to the suppression of backscattering for monolayer graphene\cite{Brihuega:2008ip}.  There has also been a topographic study of the effect of an armchair edge on the density of states in graphene on SiC which shows an oscillation along the C-C bonds that does not depend on energy\cite{Yang:2010ij}.  However, none of these studies has observed a long wavelength oscillation in the density of states controlled by the energy of the quasiparticles along with its spatial decay.  In this paper, we use scanning tunneling spectroscopy to directly probe the local density of states (LDOS) near step edges in monolayer graphene.  We observe a long wavelength oscillation of the LDOS due to scattering from the step edge.  The wavelength of the oscillation is controlled by the energy of the quasiparticles.  We also observe a faster decay ($y^{-3/2}$) of these oscillations as compared to noble metals ($y^{-1/2}$) due to the chiral nature of the quasiparticles in graphene.

The measurements were done in an ultrahigh vacuum, low-temperature STM (Omicron LT-STM) that was cooled to 4.5 K.  In order to observe the long wavelength standing waves in the density of states, a very flat surface with few scattering centers other than the step edges was needed.  Exfoliated graphene on SiO$_2$ has about a 1 nm roughness due to the roughness of the underlying oxide\cite{Ishigami:2007cl,Stolyarova:2007ku}.  In order to have exfoliated graphene laying flat on a substrate we used hexagonal boron nitride (hBN) as a substrate.  The sample was fabricated in a similar manner to previous graphene on hBN work\cite{Dean:2010jy,Xue:2011dv}. Step edges in hBN occur naturally during the sample preparation procedure.  The graphene lies flat on the hBN causing steps in the hBN to be transferred to the graphene.

\begin{figure}[h]
\includegraphics[width=0.5 \textwidth]{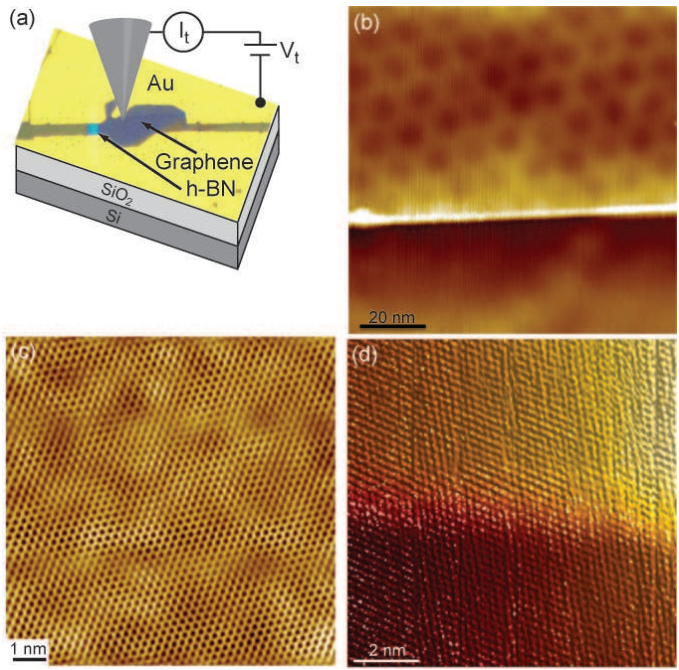}
\caption{\label{Topography} (color online) (a) Schematic illustration of the measurement setup showing the Au electrodes and graphene flake on top of a stepped hBN flake. (b) 100 nm $\times$ 100 nm topographic image of a step edge in the boron nitride beneath the graphene.  (c) Atomic resolution image of the graphene lattice on the bottom of (b) showing a $\sim$2 nm Moire pattern due to the underlying hBN.  (d) Atomic resolution image of the graphene going over the step edge showing that it is continuous.  All of the topography images were acquired with a tip voltage of -0.3 V and a tunnel current of 100 pA. }
\end{figure}

Figure \ref{Topography}(a) shows a schematic of the experimental setup.  The graphene flake is held at ground and the voltage on the tip is changed with respect to it.  The tunnel current is measured from the tip to the sample.  Figure \ref{Topography}(b) shows a 100 $\times$ 100 nm$^2$ topography image with a step edge running almost horizontally. The step height is about 0.5 nm. On the higher portion (upper portion of Fig. \ref{Topography}(b)) of the step edge, a 10 nm Moire pattern can be seen. This is due to the lattice mismatch and rotation between the graphene and the underlying hBN as was reported earlier \cite{Xue:2011dv,Decker:2011wh}.  Figure \ref{Topography}(c) shows a zoom-in on the lower portion (bottom of Fig. \ref{Topography}(b)) also showing a Moire pattern, but with a shorter $\sim$2 nm period. The different Moire patterns are a clear signature of a relative rotation between the graphene and hBN lattices. An atomically resolved image (Fig. \ref{Topography}(d)) across the step edge shows that the graphene is continuous and therefore the step edge is formed in the hBN layer. The 0.5 nm step is probably caused by an extra piece of hBN with a different lattice orientation. This different orientation leads to the two different Moire patterns on either side of the step edge.

\begin{figure*}[t]
\includegraphics[width=1 \textwidth]{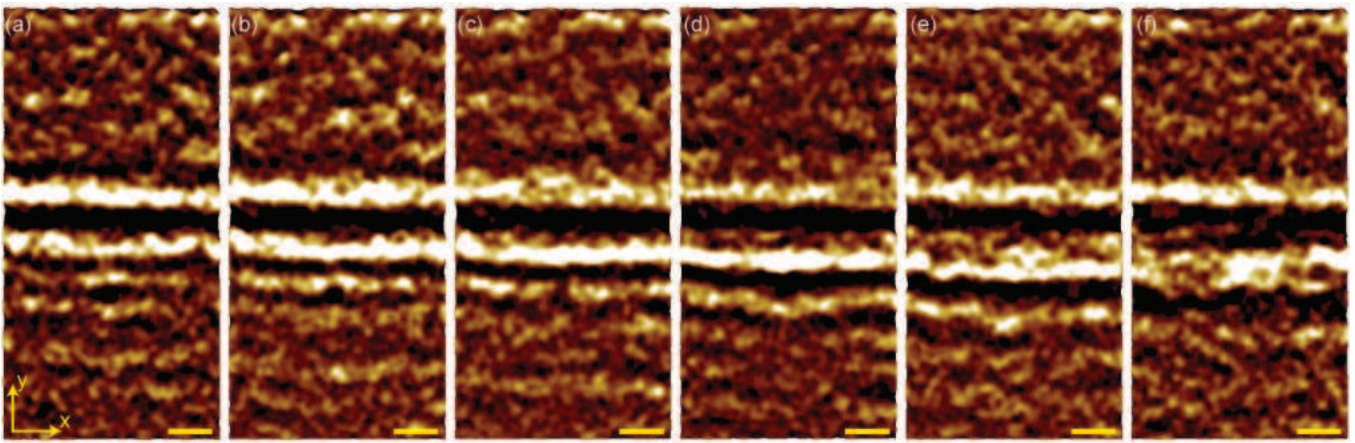}
\caption{\label{DOSImages} (color online)  (a)-(f) Images of the local density of states in graphene across a step edge running horizontally as a function of tip voltage.  The tip voltages are (a) 138 mV (b) 118 mV (c) 98 mV (d) 78 mV (e) 58 mV and (f) 38 mV.  The scale bar in all of the images is 10 nm.  For all of the images, the tip height was stabilized at a tip voltage of -0.3 V and a current of 100 pA.}
\end{figure*}

To study the influence of the step edge on the electronic properties of graphene, we have performed scanning tunneling spectroscopy (STS) measurements over the area near the step edge. In these measurements, we recorded dI/dV, which is proportional to the LDOS, as a function of position and energy.  The tip height was first stabilized with a tip voltage of -0.3 V and 100 pA.  Then the feedback circuit was turned off and a 5 mV ac voltage at 563 Hz was added to the tip voltage.  dI/dV as a function of tip voltage was recored using lockin detection in order to create images of the LDOS.  Figures \ref{DOSImages}(a)-(f) are maps of dI/dV for different tip voltages acquired in the region shown in figure \ref{Topography}(b). Two interesting features are noticed. First, LDOS oscillations in the lower portion are present and they decay moving away from the step edge. There are also LDOS oscillations in the upper portion but the signal is significantly weaker. Second, the wavelength of the oscillations increases as the tip voltage decreases approaching the Fermi energy and Dirac point.

LDOS oscillations have been previously observed on Cu\cite{Crommie:1993co} and Au\cite{Hasegawa:1993dq} surfaces. However, there is an obvious difference from our case. The wavelength of standing waves on noble metals is an order of magnitude shorter compared with that of graphene (1 nm compared to 10 nm). We will see later that this difference comes from the different band structures.  Long wavelength LDOS oscillations have been observed in other systems where the Fermi surface is not near the $\Gamma$ point\cite{McElroy:2003ed,Vonau:2005hr}.  However, the chiral nature of the quasiparticles and suppression of backscattering in graphene leads to a different decay behavior of the oscillations as compared to these other materials.  Recently, similar effects were also reported on topological insulators\cite{Anonymous:ws,Wang:2011tla} and explained by theoretical calculations\cite{Zhou:2009fz}. Since both graphene and topological insulators are Dirac systems, they share a similar linear dispersion relation and chirality of electrons.

\begin{figure}[h]
\includegraphics[width=0.48 \textwidth]{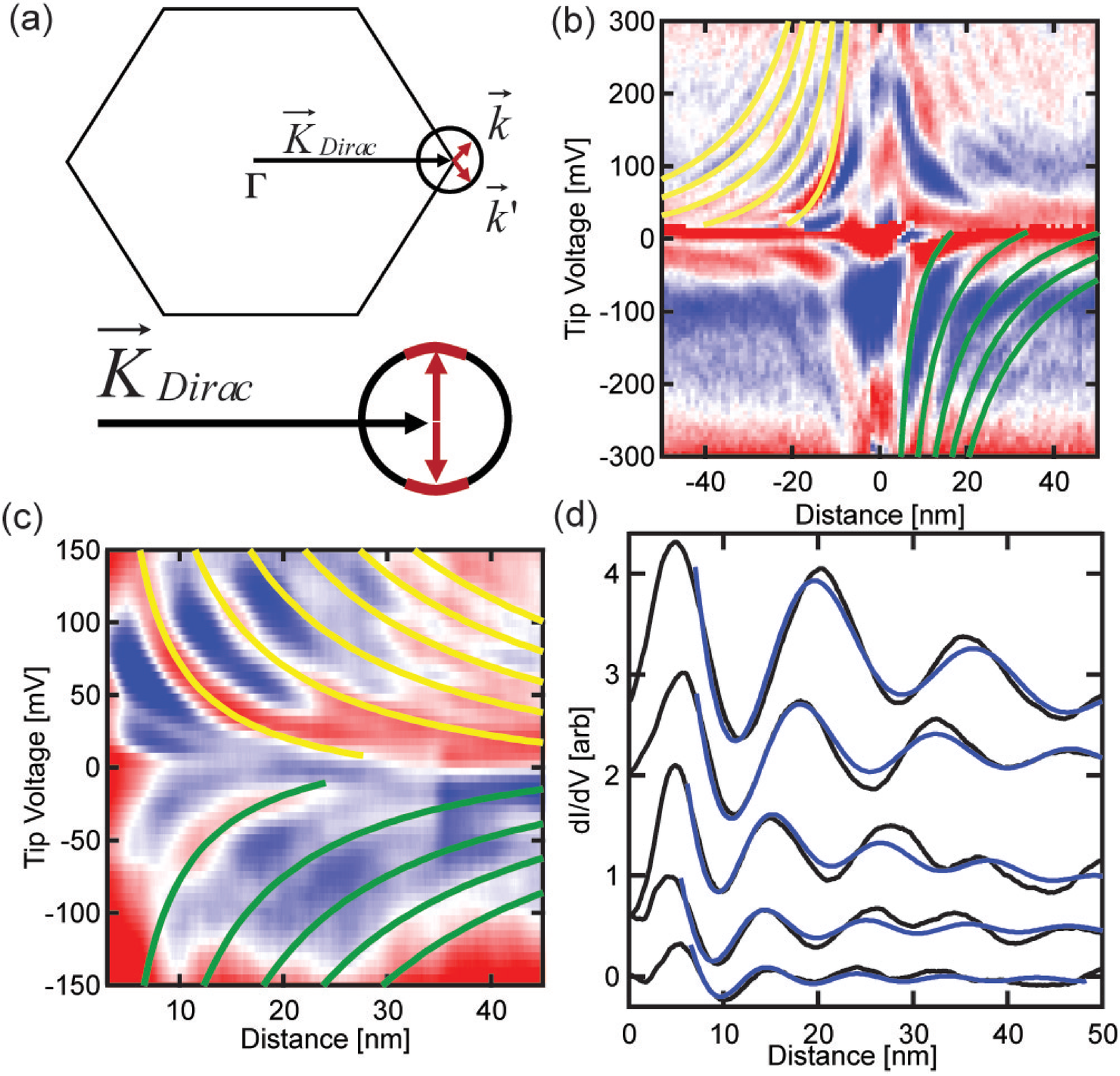}
\caption{\label{Lines} (color online) (a) Schematic diagram of the scattering process.  Quasiparticles near the Dirac point are scattered from $\vec{k}$ to $\vec{k'}$.  In this process the $x$ component of the wavevector must be conserved because the step edge is along the $x$ direction.  (b) Density of states as a function of tip voltage and position.  The data is averaged over the images in figure \ref{DOSImages}.  Blue areas are high density of states while red represents low density of states.  The colored lines represent the expected minima in the density of states from the theory calculations.  (c) Density of states as a function of tip voltage and position for a different location showing similar behavior as in (b).  (d) dI/dV as a function of position for several different tip voltages showing the decay with distance.  The black curves are the experimental data and the blue curves are from the theoretical calculations.}
\end{figure}

We adapt the theory for topological insulators\cite{Zhou:2009fz} to our case of graphene.  A step edge in graphene due to curvature can produce scalar and vector potentials which act as scatterers for electron waves\cite{Kim:2008hr}.  We define the direction along the step edge as the $x$ axis (See figure \ref{DOSImages}(a)).  Consider an incident electron with wavevector $\vec{K} = \vec{K}_{Dirac} + \vec{k}$ where $\vec{K}_{Dirac}$ is the crystal momentum at the Dirac point and $|\vec{k}|=k$ is the radius of the constant energy circle in the Dirac cone.  This electron will be partially reflected to a wavevector  $\vec{K'} = \vec{K}_{Dirac} + \vec{k'}$.  The electron will remain in the same Dirac cone as long as the step edge is not atomically sharp (the height of the graphene in Fig. \ref{Topography}(d) changes over a distance of about 2 nm) and intervalley scattering is forbidden.  A scattering barrier running in the $x$ direction conserves the $x$ component of $\vec{k}$ in the scattering process. So if $\vec{k} = \vec{k}_x + \vec{k}_y$ then $\vec{k'} = \vec{k}_x - \vec{k}_y$ in order to conserve energy and momentum (see Fig.  \ref{Lines} (a) for illustration). The incident and scattered wave interfere with each other, and give rise to a spatially modulated electron density. So far we have only considered a specific pair of wavevectors, $\vec{k}$ and $\vec{k'}$. In the actual scattering process, there are a range of wavevectors which are possible.  All of the possible incoming and scattering wavevectors $\vec{k}$ lie on a constant energy circle (CEC) determined by the tip voltage. In order to find the total change in the electron density, we must integrate the contributions from all of them, which gives, 
$$\rho(E,y) = \oint_{CEC} |\Psi_i(k_x,k_y)|^2 dk$$
where $\Psi(k_x,k_y)=\frac{\psi(k_x,k_y)+r\psi(k_x,-k_y)}{\sqrt{1+|r|^2}}$ and r is the reflection coefficient.  Since the electrons in graphene are Dirac fermions they obey the 2D Dirac equation and the wave function in momentum space near the Dirac point is given by
$$\psi(k_x,k_y) = \frac{1}{\sqrt{2}}  \begin{pmatrix} e^{-i\theta_k/2} \\ e^{i\theta_k/2} \end{pmatrix} e^{i(k_x x + k_y y)}$$
where $\theta_k = \arctan(\frac{k_x}{k_y})$.  After reflecting off the barrier, the phase factor changes to $\theta_{k}' = \pi - \theta_k$.  Then the spatially varying portion of the density of states is given by
$$\delta \rho(E,y) = \oint_{CEC} \frac{4r}{1+|r|^2}\cos(2k_y y)\sin(\theta_k) dk$$
For large distances, the integrand varies rapidly and the only points that contribute are when $k_y$ is approximately constant.  This occurs if $k_y$ is near its maximum value and hence the scattering is at nearly normal incidence (see Fig. \ref{Lines} (a)).  If the step edge is modeled as a potential barrier, then the reflection coefficient is given by $r \propto \sin(\theta_k)$ for small $\theta_k$\cite{Zhou:2009fz}.  Using the method of stationary phase, the integral can be evaluated.  Making the substitution, $E = \hbar v_F k$ where $v_F$ is the Fermi velocity, we obtain an expression for the variation in the LDOS as
\begin{equation}
\delta \rho(E,y) \propto \frac{\cos(\frac{2E y}{\hbar v_F} - 3\pi/4)}{(E y)^{3/2}} \label{DOS}
\end{equation}
This expression describes the LDOS as a decaying oscillation with minima spaced by $\pi\hbar v_F/E$. Since we only probed energies near the Dirac point, which means the radius of the CEC and hence $E$ is small, we get long wavelength LDOS oscillations compared with that from noble metals, where $E$ is usually much larger. Another interesting feature of this expression is the decay rate. In noble metals, the LDOS oscillations decay with distance from the step edge as $1/\sqrt{y}$\cite{Crommie:1993co}. However, due to the pseudo-spin degree of freedom in graphene, backscattering is forbidden. This gives a faster, $1/y^{3/2}$ decay rate to the LDOS oscillation.

Now we compare the experimental data with the theoretical prediction.  Figure \ref{Lines} (b) plots dI/dV as a function of tip voltage and distance away from a step edge.  The data is acquired by averaging, in the $x$ direction, all of the data in the two-dimensional images in figure \ref{DOSImages}, leaving a one dimensional plot of the LDOS versus distance.  There has been an overall linear slope removed in the dI/dV curves as a function of energy.  The data then shows oscillations in the density of states with blues areas having higher density of states and red areas being low.  The green and yellow lines show the minima in the density of states oscillations as predicated by equation ~(\ref{DOS}).  The only free parameter is the value of the Fermi velocity.  We find slightly different values on the bottom (negative distances) and top (positive distances) of the step.  For the yellow lines we found $v_F \approx 0.55 \times 10^6$ m/s and for the green lines we found $v_F \approx 0.75 \times 10^6$ m/s.  Figure \ref{Lines} (c) shows similar results from a different location which gives $v_F \approx 0.45 \times 10^6$ m/s for the yellow lines and $v_F \approx 0.5 \times 10^6$ m/s for the green lines.  These values are lower than other measurements of the fermi velocity in graphene by about a factor of two\cite{Li:2009eh,Zhang:2009ce,Jung:2011dk}.  The low values are confirmed by gate dependent spectroscopy measurements near the step edge which also give a reduced value of the Fermi velocity.  The reduced value of the Fermi velocity may be due to the Moire pattern, as theoretical predictions have indicated that the fermi velocity can be strongly renormalized for 2D periodic potentials which cause the Moire pattern\cite{Park:2008eg}.  As we observe two different Moire patterns on top of the step and below it, this may explain the differences in Fermi velocity observed.  The reduced Fermi velocity may also be due to strain in the graphene caused by the step edge\cite{Pereira:2009iza}.

In addition to the oscillatory nature of the LDOS, we also expect it to decay with distance from the step edge as $1/y^{3/2}$.  Figure \ref{Lines}(d) plots dI/dV as a function of distance for a series of 5 tip voltages ranging from -30 mV to -90 mV in steps of -15 mV.  These curves also show fits (in blue) of equation ~(\ref{DOS}).  The curves show the $1/y^{3/2}$ decay expected for graphene in contrast to the $1/y^{1/2}$ seen in conventional metals.  From these fits we can extract both the amplitude and wavevector of the oscillations.

\begin{figure}[h]
\includegraphics[width=0.48 \textwidth]{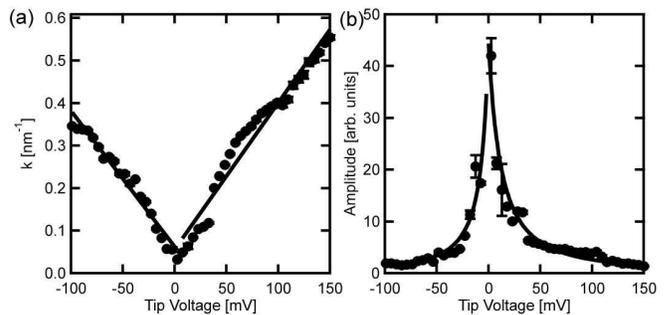}
\caption{\label{Analysis} (a) Wavevector versus tip voltage extracted from fitting the dI/dV curves in figure \ref{Lines}(c).  The solid line is a linear fit as expected for the graphene dispersion relation.  (b) Amplitude of the LDOS oscillations as a function of tip voltage (energy).  The solid lines are $1/E^{3/2}$ as expected from theory.}
\end{figure}

Figure \ref{Analysis}(a) plots the measured wavevector as a function of tip voltage for the data in figure \ref{Lines}(c).  The dI/dV curve at each energy was fit using equation ~(\ref{DOS}) to obtain the wavevector and amplitude at the given energy.  We see that the wavevector is linear in tip voltage as expected from the dispersion relation of graphene giving a value of $v_F = 0.50 \pm 0.05 \times 10^6$ m/s.  This linear relationship justifies the fits in figure \ref{Lines}(b) and (c) that assumed a linear relationship between $E$ and $k$ and used the fermi velocity as the fitting parameter.  We can also plot the amplitude of the oscillations as a function of tip voltage which is shown in figure \ref{Analysis}(b).  From equation ~(\ref{DOS}) we see that the amplitude of the oscillations is expected to decrease with increasing wavevector and hence energy.  This is what is observed in figure \ref{Analysis}(b) which shows a decrease of amplitude at larger tip voltages.  The solid line is the expected decay of the amplitude as $1/E^{3/2}$ which agrees well with the measured data.

In conclusion, we have demonstrated LDOS oscillations near step edges in graphene on hBN.  These oscillations are long wavelength and are controlled by the energy of the quasiparticles.  The decay of the oscillations is much faster than in fermi liquids and is a direct consequence of the chiral nature of the quasiparticles in graphene.
 
\begin{acknowledgments}
We are thankful for discussions with Philippe Jacquod, Aharon Kapitulnik, Leonid Levitov and Young-Woo Son. J.X. and B.J.L. acknowledge the support of NSF EECS/0925152 and NSF CAREER DMR/0953784.  J.S-Y. and P.J-H. were primarily supported by the US Department of Energy, Office of Basic Energy Sciences, Division of Materials Sciences and Engineering under Award DE-SC0001819 and partly by the 2009 US Office of Naval Research Multi University Research Initiative (MURI) on Graphene Advanced Terahertz Engineering (Gate) at MIT, Harvard and Boston University.
\end{acknowledgments}


%\bibliography{references}
\bibliography{manuscript}

\end{document}